\documentclass[aip,jcp,reprint,twocolumn]{revtex4-1}
\usepackage{times}
\usepackage{xspace}
\usepackage{graphicx}
\usepackage{color}
\usepackage{amsmath, amsthm, amssymb}
\usepackage{ulem}
\usepackage{bm}
\usepackage{tabularx}
\usepackage{hyperref}

\newcommand{\jhu}{Departments of Physics \& Astronomy and Biophysics, Johns Hopkins University, Baltimore, Maryland 21218, USA}
\newcommand{\ucsbphysics}{Department of Physics, University of California,
                          Santa Barbara, California 93106, USA}

\newcommand{\ucsbchem}{Department of Chemistry and Biochemistry,
   University of California, Santa Barbara, California 93106, USA}

\newcommand{\eq}{Eq.~}
\newcommand{\fig}{Fig.~}

\newcommand{\vb}[1]{ {\bf #1}}

\newcommand{\rb}{\ensuremath{\vb{r}}\xspace}

\newcommand{\kv}{\vb{k}\xspace}

\newcommand{\qv}{\vb{q}\xspace}

\newcommand{\lsd}{\ensuremath{L_{sd}}\xspace}

\newcommand{\Mm}{\ensuremath{\mathcal{M}}\xspace}

\begin{document}

\title{Motion of objects embedded in lipid bilayer membranes: advection and effective viscosity}
\author{Brian~A.~Camley}
\email{bcamley@jhu.edu}
\affiliation{\jhu}

\author{Frank~L.~H.~Brown}
\affiliation{\ucsbchem}
\affiliation{\ucsbphysics}

\begin{abstract}
An interfacial regularized Stokeslet scheme is presented to predict the motion of solid bodies (e.g. proteins or gel-phase domains) embedded within flowing lipid bilayer membranes.  The approach provides a numerical route to calculate velocities and angular velocities in complex flow fields that are not amenable to simple Fax\'en-like approximations.  Additionally, when applied to shearing motions, the calculations yield predictions for the effective surface viscosity of
dilute rigid body-laden membranes.  In the case of cylindrical proteins, effective viscosity calculations are compared to two prior analytical predictions from the literature. Effective viscosity predictions for a dilute suspension of rod-shaped objects in the membrane are also presented.
\end{abstract}


\maketitle

\section{Introduction}

Protein motion on the surface of lipid membranes is critical to a wide variety of biological function, from cell signaling to immune response \cite{gennis}. Understanding experimental trajectories of proteins and protein aggregates on membrane surfaces \cite{saxton97} either in vitro or in vivo requires theory that can predict the influence of protein shape and size on interactions with the surrounding lipid environment. The classic description of membrane protein diffusion is the Saffman-Delbr\"uck (SD) model \cite{saffmandelbruck75,saffman,hpw81}, which describes proteins as rigid cylinders embedded in a thin, effectively two-dimensional, fluid membrane in contact with semi-infinite bulk fluids to either side -- the water and cytosol. Because of this interesting combination of hydrodynamic flow in two and three dimensions, membranes are often referred to as ``quasi-two-dimensional" \cite{oppenheimerdiamant2009}.

The SD model predicts a verifiable dependence of diffusion coefficient on object size, membrane viscosity, and the viscosities of the surrounding bulk fluids. Many experiments on protein diffusion in membranes {\it in vitro} agree with the SD model \cite{weiss2013quantifying,ramadurai_proteins2009}, though inconsistencies have been reported by some groups \cite{urbach2006}.  Molecular dynamics simulations of lipid bilayers also validate the SD model, but it is essential to explicitly account for the small box sizes and periodic boundary conditions when comparing to theory in this case \cite{camley2015strong,venable2017lipid,vogele2016divergent,vogele2018hydrodynamics,zgorski2016toward}.  In addition, the SD model and extensions to it \cite{levine_liverpool_mackintosh_pre,levine_liverpool_mackintosh_prl} have been successfully applied far beyond the original membrane context, to the dynamics of thin layers of proteins at air-water interfaces \cite{prasadweeks2006}, soap films \cite{prasad2009two}, and suspended liquid crystal films \cite{klopp2017brownian,nguyen2010crossover}, though some interesting deviations from perfect quasi-2D behavior have also been found \cite{lee2009interfacial}. Motion of larger ordered domains on the membrane surface is also described by the SD theory \cite{cicuta2007,hormel2015two}. While deviations from SD behavior have been reported, there is no doubt that the underlying hydrodynamic picture is largely correct and should be viewed as the appropriate starting point or first-order model for studying membrane hydrodynamics.  This work assumes the validity of the SD approach and extends its predictions, via numerical calculations, to cases where analytical predictions would be difficult or impossible.

We have previously worked on extending the Saffman-Delbr\"uck model beyond its base assumptions, including the introduction of an interfacial regularized Stokeslet (RS) method to allow for numerical computations of diffusion coefficients and pair diffusion coefficients for membrane-embedded objects of arbitrary shape \cite{camley2013diffusion,noruzifar2014calculating}.  The focus of this earlier work centered around the computation
of drag and mobility coefficients, i.e. the response of an object to an applied force in an otherwise quiescent membrane, which allows for prediction of diffusion coefficients via the Einstein relation. The motion of a protein or other solid body induced by flow within the membrane was not previously considered.  This paper demonstrates how
to generalize the interfacial RS approach to compute the dynamics of force- and torque-free objects embedded in an external flow field.  Additionally, it is shown that these computations can be used to predict the effective surface viscosity of a rigid-body-laden membrane at low concentration, i.e. the intrinsic viscosity of a membrane-embedded object or the ``membrane Einstein correction".  Carrying out this calculation numerically, as opposed to analytically \cite{oppenheimerdiamant2009,henle2009effective}, opens up the possibility to consider arbitrarily shaped inclusions and
not only simple cylindrical bodies.  In particular, this work shows that linear oligomers display an increased
intrinsic viscosity relative to their monomeric counterparts, potentially allowing a characterization of membrane protein oligomerization state through determination of membrane viscosity.

\section{Regularized Stokeslets for force- and torque-free embedded objects}
\label{sec:forcefree}

\begin{figure}[htb]
 \includegraphics[width=85mm]{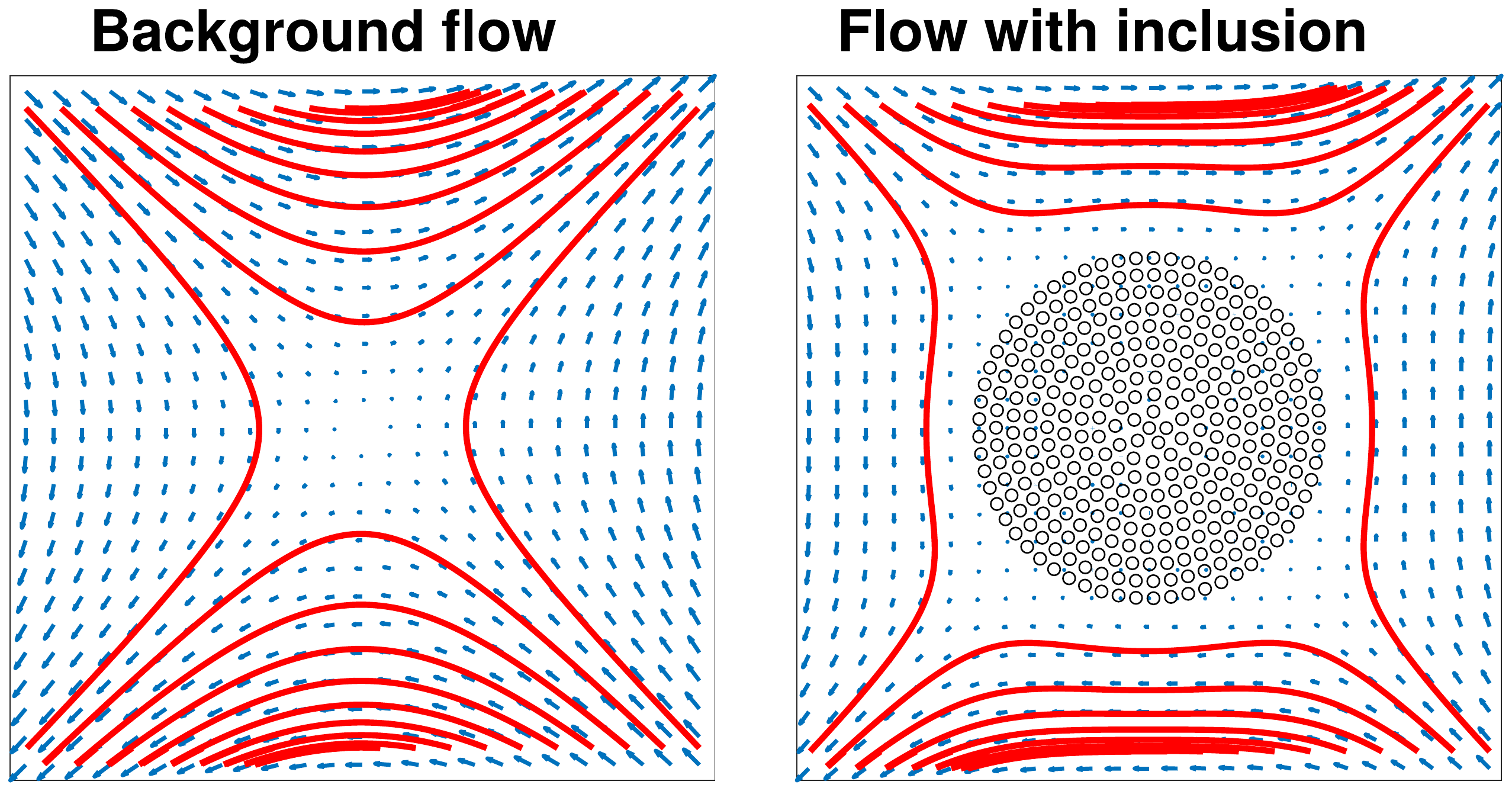}
 \caption{Effect of an embedded force- and torque-free circular object of radius $a$ on background flow $\vb{v}^\textrm{amb} = (y,x)$. Flows are shown by vector field and streamline. The object is represented by $N = 331$ blobs, with spacing $s = 0.1 a$ ($\epsilon = 0.05 a$), and  $\lsd = 100 a$; the object radius is $a = 1$. This is a coarser representation than that used for most calculations in this paper.  To the precision in the RS method ($\sim 10^{-4}$), the object remains stationary under the flow. }
 \label{fig:embedded}
\end{figure}

As for traditional three-dimensional geometries, the quasi-two-dimensional Stokes equations appropriate for the SD geometry can be solved by a Green's function approach.  Within the membrane,
\begin{equation}
v_i(\rb) = \int d\vb{r}' T^{SD}_{ij}(\rb-\rb')f_j(\rb') \label{eq:oseen}
\end{equation}
where $v_i$ are velocity components in the plane of the membrane, $T^{SD}_{ij}(\rb)$ is the membrane Oseen tensor (Green's function for velocity response to point forcing), and $f_j$ the force density in the fluid. (Here, and throughout the paper, the Einstein summation convention is used.  $i$ and $j$ run over the  $x,y$ Cartesian dimensions spanning the membrane surface.)
$T^{SD}_{ij}$ in \eq \ref{eq:oseen} is most conveniently specified by its Fourier transform \cite{lubenskygoldstein96,levinemackintosh2002}, 
\begin{equation}
\label{eq:lubensky}
T^{SD}_{ij}(\kv) = \frac{1}{\eta_m k^2 + 2 \eta_f k } \left(\delta_{ij} - \frac{k_i k_j}{k^2} \right)
\end{equation}
where $\eta_m$ is the membrane surface viscosity and $\eta_f$ the viscosity of the bulk fluid exterior to the membrane. 
(Our convention for Fourier transforms is $f(\kv) = \int d^2 r e^{-i \kv\cdot\rb} f(\rb)$, $f(\rb) = \int \frac{d^2k}{(2\pi)^2} e^{i\kv\cdot\rb} f(\kv)$. ) This Oseen tensor reduces to that of a two-dimensional fluid for lengths well below the characteristic SD length $\lsd \equiv \eta_m/2\eta_f$, and becomes similar to that of a three-dimensional fluid for lengths beyond $\lsd$ (Ref. \onlinecite{oppenheimerdiamant2009}). 

\eq \ref{eq:oseen} predicts the velocity $\vb{v}(\rb)$ of a homogeneous membrane devoid of any particles as generated
by external forcing elsewhere in the fluid $\vb{f}(\rb')$.  This formulation is not obviously useful for the description of solid bodies within the membrane.  However, the method of regularized Stokeslets (RS) \cite{cortez_fauci2005,cortez_regularized} uses the homogeneous Green's function formulation to provide numerical solutions for hydrodynamic problems that include embedded solid particles.  The essential trick is to recognize that, within creeping flow, there is no mathematical distinction between a solid body coupled to the surrounding fluid via no-slip boundaries and a ``fluid" region occupying the same space as the solid body, but constrained to undergo only rigid-body motion.  Thus, solid bodies within the RS approach
are represented as localized fluid regions subject to forcing, but constrained to translate and rotate as a solid body.
(The regularized Stokeslet approach is closely related to earlier ``shell'' methods in three dimensions \cite{bloomfield1967frictional,garcia2007improved,de2000calculation}.  In the interfacial context, related 
methods have been suggested by Levine, Liverpool, and Mackintosh \cite{levine_liverpool_mackintosh_pre,levine_liverpool_mackintosh_prl} and extended by others \cite{kuriabova2016hydrodynamic,qi2014mutual}.)

In practice, the interfacial RS scheme works as follows.  (See Refs. \onlinecite{camley2013diffusion,noruzifar2014calculating} for further details.)  The continuous force distribution $\vb{f}(\rb)$ in \eq \ref{eq:oseen} is replaced with a discrete collection of ``blobs'' with profile $\phi_\epsilon(\rb) = \frac{1}{2\pi \epsilon^2} e^{-r^2/2\epsilon^2}$.  (See Fig. \ref{fig:embedded}.) A blob centered at $\vb{r}'$, has a force distribution $\vb{f}(\rb) = \vb{g} \phi_\epsilon(\rb-\vb{r}')$ and creates the velocity response $v_i(\rb) = T^{SD}_{ij}(\rb - \rb';\epsilon) g_j$. Numerically practical formulas for $T^{SD}_{ij}(\rb-\rb';\epsilon)=\int d \vb{r}'' T_{ij}^{SD}(\vb{r}-\vb{r}'')\phi_\epsilon(\vb{r}'' - \vb{r}')$ and implementation details can be found in \cite{camley2013diffusion,noruzifar2014calculating}.

A membrane-embedded solid object is discretized into $N$ blobs centered at locations $\vb{R}_n$. External forces exerted on blob $n$ are denoted $g[\vb{R}_n]$ and the velocity response to these forces at a point $\rb$ is then
  $v_i(\rb) = \sum_n T_{ij}^{SD}(\rb-\vb{R}_n ; \epsilon) g_j[\vb{R}_n]$.
Because the Stokes equations are linear, if there is a pre-existing ambient lipid flow with velocity $\vb{v}^\textrm{amb}(\rb)$, the membrane flow at each blob becomes:
\begin{equation}
  v_i[\vb{R}_m] = v^\textrm{amb}_{i}(\vb{R}_m) + \sum_n T_{ij}^{SD}(\vb{R}_m-\vb{R}_n ; \epsilon) g_j[\vb{R}_n]; \label{eq:reg}
  \end{equation}
the second contribution indicating the influence of any external forcing on the blobs.  It is worth emphasizing that ``external forcing" includes the forces of constraint within the object that act to maintain rigid body motion in defiance of $\vb{v}^\textrm{amb}$.  These forces are imposed upon the homogeneous fluid and share the same sign convention as would a force from an external field acting on the fluid.  

To determine the motion of a solid body or particle carried under the influence of $\vb{v}^\textrm{amb}(\rb)$, but absent any additional external forcing, requires the determination of $g_j$s that 1) create only rigid body motion within the body,  i.e.
\begin{equation}
\vb{v}\left[\vb{R}_m\right] = \vb{U} + \boldsymbol\Omega \times \vb{R}_m
\label{eq:rigid_body}
\end{equation}
with $\vb{U}$ and $\boldsymbol\Omega$ specifying the particle's velocity and angular velocity; 2) combine to yield vanishing total external force and torque on the particle, i.e. 
\begin{eqnarray}
\sum_n \vb{g}\left[\vb{R}_n\right] = 0, \; \; \;  
\sum_n \vb{R_n} \times \vb{g}[\vb{R}_n] = 0
\label{eq:vanishing}
\end{eqnarray}
(assuming the centroid of the blobs is at the origin); and 
3) satisfy Eq. \ref{eq:reg}.  
Eqs. \ref{eq:reg} - \ref{eq:vanishing} represent $4N+3$ equations in $4N+3$ unknowns
and could, in principle, be naively solved to yield $\vb{v}\left[\vb{R}_m\right]$ ($2N$ scalars), $\vb{g}\left[\vb{R}_n\right]$ ($2N$ scalars),
$\vb{U}$ (2 scalars) and $\boldsymbol\Omega$ (1 scalar). In practice, it is convenient
to equate Eqs. \ref{eq:reg} and \ref{eq:rigid_body} by removing the blob velocities as independent variables.
The remaining $2N+3$ equations in $2N+3$ unknowns can be solved to yield  $\vb{g}\left[\vb{R}_n\right]$,
$\vb{U}$ and $\boldsymbol\Omega$. Then, $\vb{v}(\rb)$ for any point $\rb$ may be directly computed from \eq \ref{eq:rigid_body} or \eq \ref{eq:reg}. 
It is convenient to explicitly use the
linearity of Eq. \ref{eq:reg} to simplify this solution.

First, a set of forces $g_j[\vb{R}_n]$ is determined from the superposition of four individual problems:
\begin{align}
  -v^\textrm{amb}_{i}(\vb{R}_m) &= \sum_n T_{ij}^{SD}(\vb{R}_m-\vb{R}_n ; \epsilon) g_j^\textrm{cancel}[\vb{R}_n] \label{eq:cancel}\\
  \hat{\vb{x}} &= \sum_n T_{ij}^{SD}(\vb{R}_m-\vb{R}_n ; \epsilon) g_j^\textrm{x-trans}[\vb{R}_n] \label{eq:xtrans}\\
  \hat{\vb{y}} &= \sum_n T_{ij}^{SD}(\vb{R}_m-\vb{R}_n ; \epsilon) g_j^\textrm{y-trans}[\vb{R}_n] \label{eq:ytrans}\\
  \hat{\vb{z}}\times\vb{R}_m &= \sum_n T_{ij}^{SD}(\vb{R}_m-\vb{R}_n ; \epsilon) g_j^\textrm{rot}[\vb{R}_n] \label{eq:rot}  
\end{align}
Here, $\vb{g}^\textrm{cancel}$ are the forces required to cancel the background flow, resulting in an object that is not moving at all, $\vb{g}^\textrm{x-trans}$ and $\vb{g}^\textrm{y-trans}$ generate uniform $x$ and $y$ translation with unit velocity, and $\vb{g}^\textrm{rot}$ creates rotation with unit angular velocity.  The blob forces for each individual problem are determined via the GMRES method, as in Ref. \onlinecite{camley2013diffusion}.  By combining Eqs. \ref{eq:cancel} - \ref{eq:rot} and comparing with \eq \ref{eq:reg}, it is clear that $\vb{g} = \vb{g}^\textrm{cancel}+U_x \vb{g}^\textrm{x-trans}+U_y\vb{g}^\textrm{y-trans}+\Omega\vb{g}^\textrm{rot}$  creates a flow with $v\left[\vb{R}_m\right] = \vb{U} + \boldsymbol\Omega \times \vb{R}_m$, i.e. perfect rigid body motion. Requiring total force and torque to vanish, 
(Eq. \ref{eq:vanishing}) yields
\begin{widetext}
\begin{equation}
  \left(\begin{array}{ccc} \sum_n g_x^\textrm{x-trans} & \sum_n g_x^\textrm{y-trans} & \sum_n g_x^\textrm{rot} \\
    \sum_n g_y^\textrm{x-trans} & \sum_n g_y^\textrm{y-trans} & \sum_n g_y^\textrm{rot} \\
    \sum_n \vb{R}_n \times \vb{g}^\textrm{x-trans} & \sum_n \vb{R}_n \times \vb{g}^\textrm{y-trans} & \sum_n \vb{R}_n \times \vb{g}^\textrm{rot}
  \end{array}\right)
  \left(\begin{array}{c}
    U_x \\ U_y \\ \Omega \end{array}\right)
    =\left(\begin{array}{c}-\sum_n g_x^\textrm{cancel}[\vb{R}_n] \\
    -\sum_n g_y^\textrm{cancel}[\vb{R}_n] \\
    -\sum_n \left(\vb{R}_n\times \vb{g}^\textrm{cancel}[\vb{R}_n]\right)_z
    \end{array}\right) \label{eq:linearsystem}.
\end{equation}
\end{widetext}
This 3x3 linear system can easily be solved in order to find the values of $\Omega$ and $\vb{U}$ required to make the object force-free. (\eq \ref{eq:linearsystem} may also be easily generalized in order to specify constraints on total force and torque, if it is necessary to consider a particle with an applied force acting on it.)

\begin{figure*}[ht!]
 \includegraphics[width=170mm]{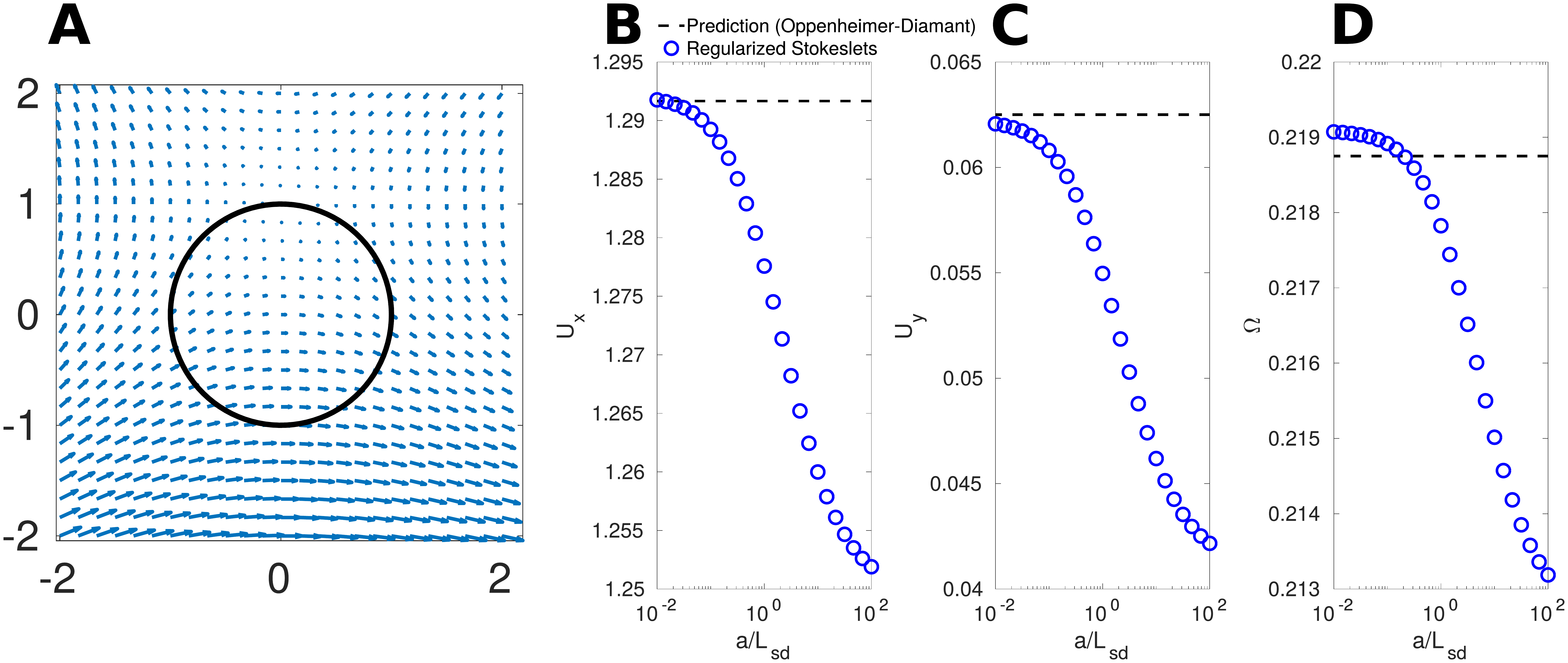}
 \caption{Computation of object motion in external flow field $\vb{v}^\textrm{amb} = \left(-(y-1)+\frac{1}{8}(y-1)^2-\frac{1}{24}(y-1)^3,-x+\frac{1}{8}x^2+\frac{1}{24}x^3\right)$. In this calculation, the size of the object is kept fixed (radius $a = 1$) but the Saffman-Delbr\"uck length is varied. {\bf A}: Illustration of flow field and location and size of object; only the background flow field is shown, not the effect of the embedded object. {\bf B}: $U_x$ (unitless) does not strongly deviate from the prediction of \eq \ref{eq:faxenvsimple}; this occurs because the dominating term is that $U_x \approx v_x^\textrm{amb}(0)$. {\bf C}: Because $v_y^\textrm{amb}(0) = 0$, $U_y$ depends strongly on the local change in $\vb{v}^\textrm{amb}$; this term shows a slightly stronger dependence on the Saffman-Delbr\"uck length. {\bf D}: Prediction of angular velocity. Even in the limit of $a \ll \lsd$, a small ($<0.2\%$) error remains. Relatively weak deviations occur for $a \gg \lsd$. Results are extrapolated from a range of spacings, $s = 0.05a, 0.1a,0.15a,0.2 a$ with $\epsilon = s/2$.}
 \label{fig:faxensyst}
\end{figure*}

As a simple test case, consider a force- and torque-free cylinder localized at the origin in a simple extensional flow $\vb{v}^\textrm{amb} = (y,x)$. By symmetry, the velocity and angular velocity of the particle are expected to be zero. This test case is displayed in \fig \ref{fig:embedded}; as expected  $U_x$, $U_y$, and $\Omega$ are small (all with absolute value $<10^{-4}$, with signs that can change depending on the spacing and specific details of the blobs chosen.)  Throughout this paper, unitless velocities will generally be reported; because of the linearity of the Stokes equations, the overall scale of the velocity is only important in setting the absolute value of the force and torque. 
The flow in \fig \ref{fig:embedded} results from a relatively rough discretization (only 331 blobs and $\epsilon = 0.05 a$, where $a$ is the particle radius). For detailed predictions, extrapolation to infinite blobs and zero spacing must
be carried out \cite{camley2013diffusion,bloomfield1967frictional,garcia2007improved,de2000calculation}.  In this work,
final extrapolated results are calculated by fixing $\epsilon$ to half of the characteristic spacing of the discretization while reducing the spacing between blobs as specified in particular examples. 

\section{Advection of membrane-embedded objects: beyond the simplest Fax\'en relationships}
\label{sec:faxen}

If a force-free sphere of radius $a$ is embedded in a three-dimensional ambient flow field $\vb{v}^\textrm{amb}$ at position $\vb{r}$, its velocity is given by the Fax\'en relationship, $\vb{U} = \vb{v}^\textrm{amb}(\rb) + \frac{1}{6} a^2 \nabla^2 \vb{v}^\textrm{amb}(\rb)$, with a similar result for the angular velocity \cite{kimkarrila}. Because of the finite size of the sphere, it does not track exactly the imposed flow at its center of mass, but rather includes an average of the flow over its spatial envelope. For membranes, Oppenheimer and Diamant derived approximate Fax\'en relationships \cite{oppenheimerdiamant2009}; for force- and torque-free cylindrical particles, these are
\begin{align}\label{eq:faxenvsimple}
  \vb{U} &\approx \left[1+\frac{1}{4} a^2 \nabla^2 \right]\vb{v}^\textrm{amb}(\rb) \\
  \boldsymbol\Omega &\approx \frac{1}{2}\left[1+ \frac{1}{8}a^2 \nabla^2 \right] \nabla \times \vb{v}^\textrm{amb}(\rb) \label{eq:faxenomegasimple}
  \end{align}
where the approximation indicates that these results only hold in the limit of $a \ll \lsd$, and could have higher-order corrections proportional to $a^4\nabla^4\vb{v}^\textrm{amb}$.

Eqs. \ref{eq:faxenvsimple} and \ref{eq:faxenomegasimple} are valid in the appropriate limit, but the interfacial RS method can capture interesting systematic deviations from these predictions. In \fig \ref{fig:faxensyst},  results for a cylindrical object in an arbitrarily chosen background field $\vb{v}^\textrm{amb}$ are presented. When $\lsd$ is decreased, so that the assumptions of Oppenheimer and Diamant no longer hold, there can be deviations from Eqs. \ref{eq:faxenvsimple}-\ref{eq:faxenomegasimple}; these effects are generally small when the first term of \eq \ref{eq:faxenvsimple} is large (as in \fig \ref{fig:faxensyst}B), but can be relatively large ($\sim 30\%$) when the leading order term is small, e.g. if the mean velocity at the object center of mass is small (\fig \ref{fig:faxensyst}C). Corrections to the angular velocity appear to be very small (\fig \ref{fig:faxensyst}D). There is a small ($<0.2\%$) deviation from the analytical results of Ref. \onlinecite{oppenheimerdiamant2009} in \fig \ref{fig:faxensyst}D, even in the appropriate limit.  The effect is exaggerated on the plot due to the scale of the $\Omega$ axis; a numerical error of $0.2\%$ is entirely consistent with the accuracy of the interfacial RS approach \cite{camley2013diffusion}. However, this error can be large in a relative sense as the error does not become smaller when the predicted angular velocity approaches zero.  It is important to be aware of the limitations of the numerical methodology -- we do not expect the RS method to produce values that are accurate to below $10^{-3}$ of the typical velocity scale without significantly more refinement than we perform here.

\begin{figure}[htb]
 \includegraphics[width=85mm]{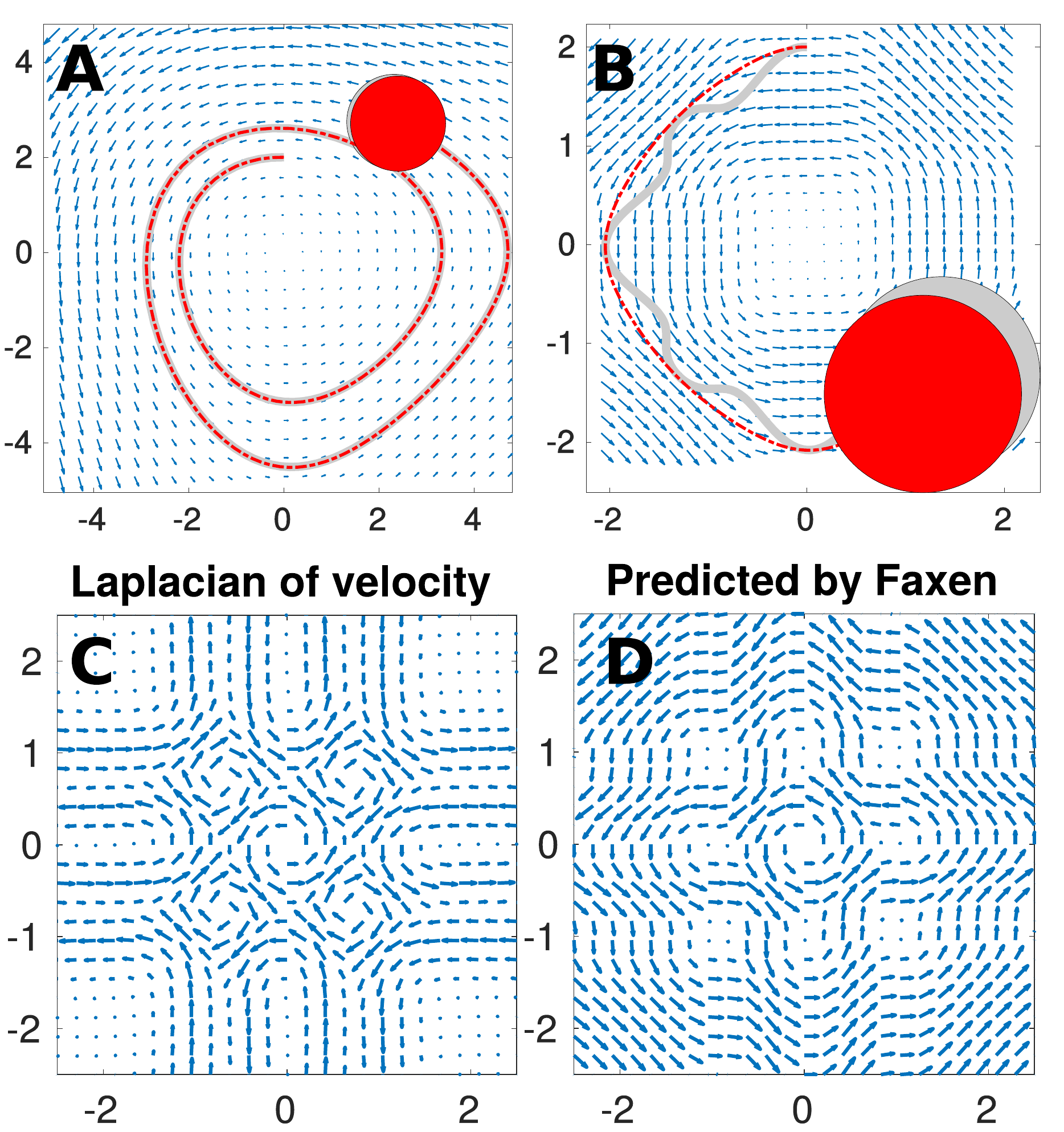}
 \caption{In both {\bf A} and {\bf B}, the motion of a membrane-embedded object in a background flow is computed by the approximate Fax\'en relationship, \eq \ref{eq:faxenvsimple} (thick gray line) and by the interfacial RS approach (red dash-dot line). The gray and red circles indicate the final position of the objects given by the Fax\'en relationship and RS, respectively. In both panels, $\lsd = 100 a$, and the axis labels are in units of the object radius $a$, and with blob spacing $s = 0.1a$ and $\epsilon = 0.05 a$ fixed. The background flow fields are $\vb{v}^\textrm{amb} = \left(-y-\frac{1}{8}y^2,x-\frac{1}{8}x^2\right)$ ({\bf A}) and $\vb{v}^\textrm{amb} = \left(-\frac{y^3}{1+y^4},\frac{x^3}{1+x^4}\right)$ ({\bf B}). Time step used is $\Delta t = 0.1$. Panels {\bf C} and {\bf D} help explain why the Fax\'en approximation fails for the flow field of Panel {\bf B}. {\bf C} shows the Laplacian of $\vb{v}^\textrm{amb}$ for Panel {\bf B}, while {\bf D} shows the Fax\'en prediction's flow field, $\left[1 + \frac{a^2}{4} \nabla^2 \right]\vb{v}^\textrm{amb}$, leading to the unphysical oscillatory trajectory displayed in panel {\bf B}.}
 \label{fig:faxenintegrated}
\end{figure}

The results of \fig \ref{fig:faxensyst} show the predicted velocity of a membrane-embedded object at a single point in space and time, demonstrating some deviations from \eq \ref{eq:faxenvsimple}. It is also possible to integrate the motion of an object through time, predicting the particle's trajectory when embedded in a stationary external lipid flow.  Doing so can reveal these deviations more dramatically. (This is a purely hydrodynamic calculation, neglecting thermal motion.) In \fig \ref{fig:faxenintegrated}, motion is tracked by computing the velocity of an object embedded in a lipid flow, and then evolving the position of the center of mass of that object by $\vb{r}_\textrm{cm}(t+\Delta t) = \vb{r}_\textrm{cm}(t) + \vb{U} \Delta t$.  The velocity $\vb{U}$ is then updated for the object's new location and the procedure is repeated and iterated. The resulting trajectories are shown in \fig \ref{fig:faxenintegrated}; similar trajectories can be computed for comparison purposes directly from \eq \ref{eq:faxenvsimple}. In the limit of a smooth flow field and $a \ll \lsd$, the regularized Stokeslet method and \eq \ref{eq:faxenvsimple} agree well (\fig \ref{fig:faxenintegrated}A). However, for a more rapidly varying background flow field, the simple Fax\'en relationship predicts a trajectory that is significantly more oscillatory than the one found by the regularized Stokeslet method (\fig \ref{fig:faxenintegrated}B). This occurs because the Fax\'en relationship has neglected higher-order derivatives, and when the background flow field develops structure on the scale of $a$, the truncation is invalid. Although it is not immediately apparent in \fig \ref{fig:faxenintegrated}B, the wiggles in the Fax\'en trajectory do result from the Laplacian term, as is clear from  \fig \ref{fig:faxenintegrated}CD.

In both \fig \ref{fig:faxensyst} and \fig \ref{fig:faxenintegrated}, comparing between the Fax\'en relationships and the interfacial RS method, it is important to choose flow fields with a sufficient number of nonzero derivatives. For instance, $\vb{v}^\textrm{amb} = (-y,x)$ would misleadingly give the impression that there are no significant deviations from \eq \ref{eq:faxenvsimple}, because $\nabla^2 \vb{v}^\textrm{amb} = 0$, and the particle is simply advected along the streamlines.

\section{Computing effective viscosity}
\label{sec:effective}
In Ref. \onlinecite{oppenheimerdiamant2009}, Oppenheimer and Diamant describe how to determine the effective viscosity of a dilute solid-body-laden membrane.  Their approach is similar in spirit to the calculations underlying Eqs. 
\ref{eq:faxenvsimple} and \ref{eq:faxenomegasimple} and is subject to the same assumptions, namely that
the solid bodies are cylindrical and that the object radius is much smaller than $\lsd$.  The interfacial RS 
approach may be used to relax these assumptions, extending the original Oppenheimer Diamant calculation to determine
the the effective viscosity for a dilute suspension of arbitrarily shaped and sized bodies within the membrane.

Following Ref. \onlinecite{oppenheimerdiamant2009},  the basic idea for this calculation is to subject the membrane to a steady point force at the origin that generates an ambient flow field determined by $T^{SD}(\vb{r})$, and observe how the presence of an ensemble of solid bodies within the lipid bilayer distorts the average long-distance response of the membrane away from the ambient flow.  If a rigid body is embedded in a membrane with a background flow $\vb{v}^\textrm{amb}(\rb)$, the body will follow the flow and rotate in response to the local velocity and circulation, but the rigid particle can not deform in response to a local shear.  The fact that the particle cannot deform means that there will be a net force dipole (and higher order multipoles as well) on the particle and a compensating force dipole (of opposite sign) imparted by the particle back into the membrane.  Because of the linearity of the Stokes equations, this force dipole must be linearly proportional to the background velocity. If the background flow is sufficiently smooth the force dipole must be proportional to the appropriate gradients of the flow in the absence of the object\cite{oppenheimerdiamant2009},
\begin{equation}
S_{ij} \approx -\alpha \eta_m A_p \left( \partial_i v_{j}^\textrm{amb}+\partial_j v_{i}^\textrm{amb} \right) \label{eq:stressfaxen}
\end{equation}
where the force dipole $S_{ij} = \frac{1}{2} \int d^2 r \left[ r_i f_j(\rb) + r_j f_i(\rb) \right] $, and $\alpha$ is a unitless number.  {Dimensional analysis indicates that $\alpha$ depends only on the object shape and $\sqrt{A_p}/L_{SD}$, where $A_p$ is the area of the object.} (Note the sign difference in \eq \ref{eq:stressfaxen} from Ref. \onlinecite{oppenheimerdiamant2009}.  The convention here is that $S_{ij}$ is the force dipole imparted by the particle on the membrane in response to the hydrodynamic dipole, $S_{hy}=-S$, acting on the particle.) This relation assumes that the background flow is slowly-varying on the length scale $a$ of the embedded object, so higher-order terms, e.g. $a^2 \nabla^2 \vb{v}^\textrm{amb}$, can be neglected. Ref. \onlinecite{oppenheimerdiamant2009} finds $\alpha = 2$ for a circular object with radius $a \ll \lsd$.  \eq \ref{eq:stressfaxen} will only hold for an {\it isotropic} object. If the object is anisotropic, \eq \ref{eq:stressfaxen} is expected to hold only after averaging over particle orientation.

The value of $\alpha$ in \eq \ref{eq:stressfaxen} determines the effective viscosity of a membrane with dilutely embedded objects. In the Appendix, following Ref. \onlinecite{oppenheimerdiamant2009}, it is shown that a membrane with area fraction $\phi \ll 1$ of solid inclusions responds, on average, as if it were a pure membrane with effective viscosity
\begin{equation}
\eta_m^{\textrm{eff}} = \eta_m ( 1 + \alpha \phi ) \label{eq:effective}
\end{equation}
In other words, $\alpha=\lim_{\phi\to0} \frac{\eta_m^{\textrm{eff}} -\eta_m}{\eta_m \phi}$ is the ``intrinsic membrane viscosity" for the embedded objects within the membrane. {Eq. \ref{eq:effective} is valid and $\alpha$ is meaningful only in the dilute limit of small $\phi$.  Simulations suggest that Eq. \ref{eq:effective} may be reasonable up to $\phi \approx 0.1$. \cite{camley2014fluctuating}}

\begin{figure*}[htb]
 \includegraphics[width=120mm]{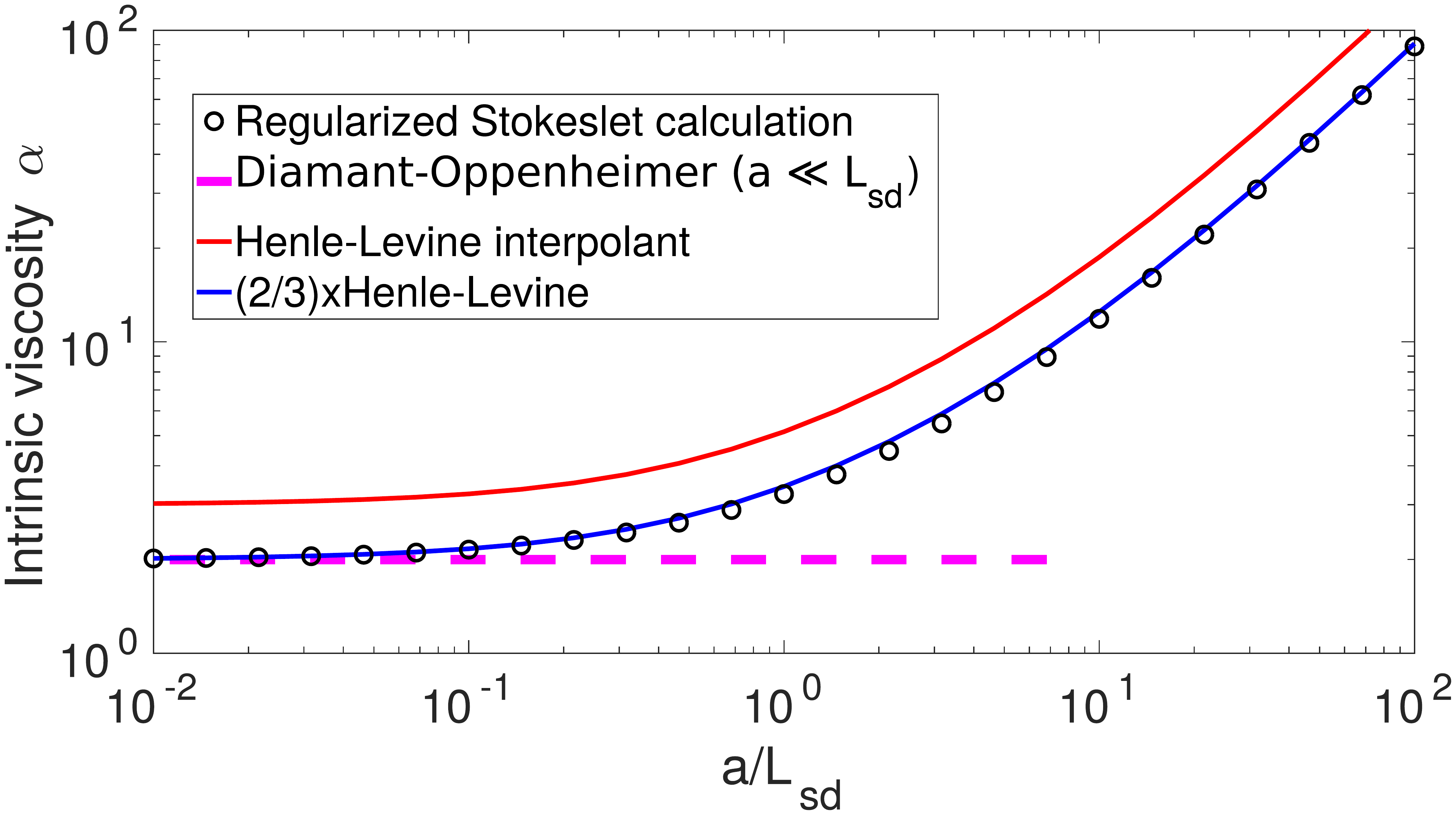}
 \caption{Comparison between numerical RS results and literature predictions for suspensions of cylindrical particles (with radius $a$) in a membrane. The numerical results match Ref. \onlinecite{oppenheimerdiamant2009} in the appropriate limit of $a \ll \lsd$, but disagree with those of Ref. \onlinecite{henle2009effective}, which defines the effective viscosity differently. However, simply rescaling the results of Ref. \onlinecite{henle2009effective} by a factor of $2/3$ gives good agreement over all $\lsd$ values (a maximum disagreement of about $7\%$). The solid lines are calculated using the interpolating function \cite{henle2009effective}: $\alpha \approx f(a/\lsd)$, with $f(z) = 12\pi^{-2}\ln(1+z) + \pi^{-2}\left(1+z\right)^{-1}\left[3 \pi^2+(3\pi^2+8\pi-12)z + 4\pi z^2\right]$. It is possible that the disagreement between RS numerics and the rescaled full predictions of Ref. \onlinecite{henle2009effective} may be significantly less than $7\%$, as Ref. \onlinecite{henle2009effective} reports full numerical results that fall slightly below the interpolant curve, similar to what is seen for the RS numerics. Reported RS results are extrapolated from a range of spacings $s = 0.05a, 0.1a, 0.15a,\cdots 0.4a$ with $\epsilon=s/2$.}
 \label{fig:compare}
\end{figure*}

To compute $\alpha$ for cylindrical particles, interfacial RS calculations are performed.  The particle is placed 
in a background flow $\vb{v}^\textrm{amb}(\vb{r}) = (x,-y)$ and the force-free $g_i[\vb{R}_n]$'s are
determined as in Sec. \ref{sec:forcefree}.  $\vb{v}^\textrm{amb}(\vb{r}) = (x,-y)$ is a convenient form to choose, because it involves a pure shearing motion and it is impossible for higher-derivative terms to complicate Eq. \ref{eq:stressfaxen}.
The stresslet can then be computed as $S_{ij} = \frac{1}{2} \sum_n \left(R_{n,i} g_j[\vb{R}_n] + R_{n,j} g_i[\vb{R}_n] \right)$ and compared to the predictions of 
\eq \ref{eq:stressfaxen}.  Given the choice of ambient flow, $S_{xx} = -S_{yy} = -2 \alpha \eta_m  A_p$ and $S_{xy} = S_{yx} = 0$ and $\alpha$ follows immediately.  In practice, $S_{ij}$ is determined at many different resolutions, and the results are linearly interpolated to estimate $S_{ij}$ as $\epsilon \to 0$. These extrapolated results are presented in \fig \ref{fig:compare}.

Consistent with the asymptotic prediction of Oppenheimer and Diamant, the numerical predictions return $\alpha \to 2$ as $a \ll \lsd$ (\fig \ref{fig:compare}). This contrasts with the result of Henle and Levine \cite{henle2009effective}, who, using a different definition of effective viscosity predict that $\alpha \to 3$ as $a \ll \lsd$ (\fig \ref{fig:compare}). The present work employs a definition of effective viscosity identical to that of Oppenheimer and Diamant, so it may not be surprising to find agreement with their results. Beyond the limit $a \ll \lsd$ treated analytically by Ref. \onlinecite{oppenheimerdiamant2009},  $\alpha$ increases, as predicted qualitatively by Henle and Levine \cite{henle2009effective}. In fact, simply multiplying Henle and Levine's results by a factor of $2/3$, to force agreement at small $a/\lsd$, yields excellent agreement over the full range of $\lsd$ values (\fig \ref{fig:compare}).  The
origin of the factor of $2/3$ is not understood.

{
The linear growth of $\alpha$ at large values of $a/L_{SD}$ (see Fig. \ref{fig:compare}) reflects the fact that solid objects embedded within the membrane influence the hydrodynamics of the membrane system, even when the bare membrane viscosity becomes vanishingly small; their presence affects the flow of the surrounding water \cite{hpw81,henle2009effective}. This point was also noted in the original work by Henle and Levine \cite{henle2009effective}. For the effect of the objects to remain finite in this limit, $\alpha$ must increase linearly in $a/L_{sd}$, to overcome the explicit factors of $\eta_m$ in Eqs.
 \ref{eq:stressfaxen} and \ref{eq:effective}.     
Alternatively, when $a\gg L_{SD}$, it is possible to formulate the influence of membrane-embedded particles in terms of an effective bulk viscosity for the fluid surrounding the membrane \cite{henle2009effective}; the bulk intrinsic viscosity in this formulation saturates to a constant when $a \gg L_{SD}$.  The unbounded growth of $\alpha$ in Fig. \ref{fig:compare} is not the indication of a physical divergence, but simply reflects the traditional conventions used in Eqs. \ref{eq:stressfaxen} and \ref{eq:effective}, where the leading effect of rigid inclusions is expressed as being proportional to the \textit{product} $\eta_m \cdot \alpha$. }

The interfacial RS approach can also determine the intrinsic viscosity of arbitrarily-shaped objects, if one additional step is added to the numerics. As noted above,  \eq \ref{eq:stressfaxen} should only be expected to hold for isotropic particles. Indeed, when $S_{ij}$ for an anisotropic object is calculated, \eq \ref{eq:stressfaxen} can be violated.  For example, under the flow $\vb{v}^\textrm{amb} = (x,-y)$, it is no longer the case that $S_{xx} = -S_{yy}$. However, if $S_{ij}$ is determined for different orientations $\theta$ of an anisotropic object, and these orientations are averaged over as $\overline{S}_{ij} = \frac{1}{2\pi}\int_0^{2\pi} d\theta S_{ij}(\theta)$, then \eq \ref{eq:stressfaxen} holds for $\overline{S}_{ij}$. Appendix \ref{app:viscosity} shows that it is this orientationally-averaged stresslet that reports the intrinsic viscosity, assuming object orientations are uniformly distributed. This averaging increases the cost of calculation, as individual RS calculations must be performed at each orientation. (We present an alternate approach which can compute this averaging analytically, significantly reducing computational cost, in Appendix \ref{app:gr}.)

To summarize, the algorithm for computation of effective viscosity is:
\begin{enumerate}
\item Set up a discretization for particle shape with a particular spacing $s$ and regularization scale $\epsilon = s/2$, with the particle centroid at the origin. \label{list:setup}
\item Solve for the rigid-body response of the particle in the external flow field $\vb{v}^\textrm{amb}(\vb{r}) = (x,-y)$ as in Section \ref{sec:forcefree}, keeping a constraint of zero net force and zero net torque on the object. This solution provides a set of forces exerted on the blobs representing the object, $\vb{g}[\vb{R}_n]$.
\item Given these forces, compute the stresslet $S_{ij}(\epsilon) = \frac{1}{2} \sum_n \left(R_{n,i} g_j[\vb{R}_n] + R_{n,j} g_i[\vb{R}_n] \right)$. \label{list:Sij}
\item Repeat steps \ref{list:setup}-\ref{list:Sij}, after having rotated the particle around the origin. Average over many rotations to approximate $\overline{S}_{ij}(s) = \frac{1}{2\pi}\int d\theta S_{ij}(s,\theta)$. (This step can be skipped for rotationally symmetric objects). \label{list:rotate}
\item Repeat steps \ref{list:setup}-\ref{list:rotate} for increasingly refined spacings $s$ (keeping $\epsilon = s/2$), and then use linear extrapolation to estimate $\overline{S}_{ij}(s \to 0)$.
\item The intrinsic viscosity $\alpha$ is given by $\alpha = \overline{S}_{yy}(s\to0) \times \left[ 2 A_p \eta_m \right]^{-1}$. 
\end{enumerate}

\begin{figure}[htb]
 \includegraphics[width=85mm]{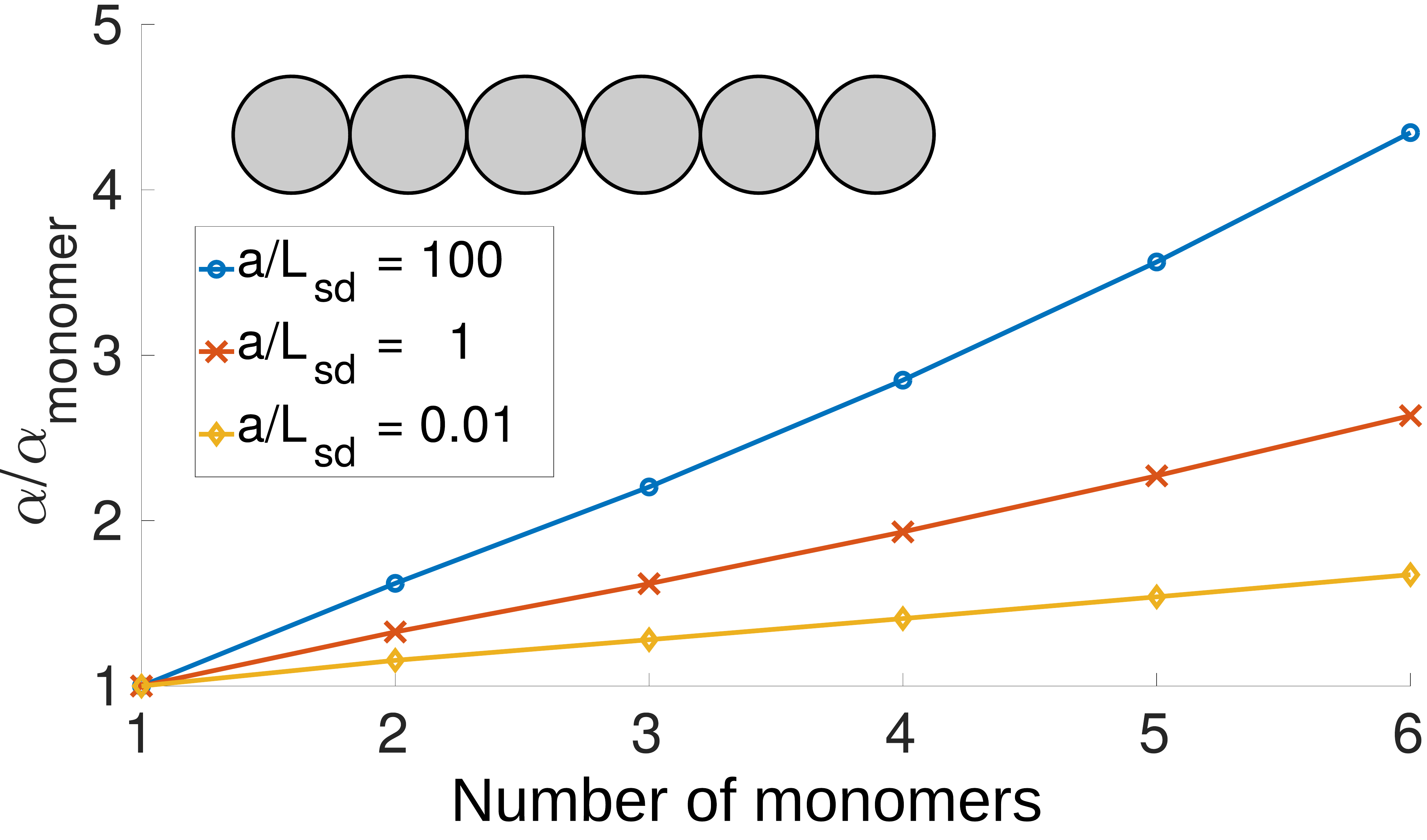}
 \caption{Intrinsic viscosity of {rigid} linear oligomers of circular objects of radius $a$. $\alpha$ is plotted as a function of the number of monomers in the oligomer. $\alpha$ increases as the number of monomers increases, and increases more quickly (relative to its original value) for objects where $a/\lsd$ is larger. This simulation is computed from extrapolation over spacings $s = 0.05 a, 0.08a, 0.11a, 0.14a$ and averaging over ten equally spaced rotations.}
 \label{fig:oligo}
\end{figure}

As a further application of this approach, the intrinsic viscosity of a suspension of {rigid} linear oligomers of circular monomers has been calculated. The intrinsic viscosity increases as a function of the length of the chain (\fig \ref{fig:oligo}).
Stated differently, the effective viscosity of the membrane is expected to increase when the oligomerization state is increased while
maintaining a fixed area fraction of embedded objects; the linear assemblies have a larger effect on viscosity than if they are broken apart.  Similar results for the intrinsic viscosity are found for chains of three-dimensional spheres in bulk fluid \cite{garcia2007improved}. Other papers have also addressed the hydrodynamic drag on extended bodies in membranes \cite{levine_liverpool_mackintosh_pre,levine_liverpool_mackintosh_prl,fischer_rods}, but not the intrinsic viscosity arising from them.

It is worth remarking that intrinsic viscosities are always defined in the limit of small area fractions $\phi \to 0$. 
The present work makes no attempt to quantify the range of $\phi$ values over which the present calculations may be reliable.  In particular, it could be the case that longer oligomers show deviations from theoretical predictions
earlier (i.e. at smaller $\phi$ values) than shorter oligomers. Previous work suggests that the linear correction works surprisingly well even to sizable area fractions for compact particles \cite{camley2014fluctuating}, but possible steric or hydrodynamic interactions may become  more relevant for elongated particles, and it is not obvious whether the linear correction to viscosity would be preserved for such high area fractions.

The calculations presented in this section closely parallel the analytical approach of Oppenheimer and Diamant \cite{oppenheimerdiamant2009}. It is also possible to compute the membrane effective viscosity via the grand resistance matrix formalism \cite{kimkarrila} for solid bodies within the membrane. This rephrasing of the problem more closely resembles the mobility/diffusion calculations in Refs. \onlinecite{camley2013diffusion,noruzifar2014calculating} and is presented in Appendix \ref{app:gr}. One advantage of phrasing our central problem in terms of resistance tensors is that the orientational averaging may be carried out analytically, leading to an order-of-magnitude speedup in computation.

\section{Discussion and Conclusions}

This paper extends the interfacial regularized Stokeslet scheme\cite{camley2013diffusion,noruzifar2014calculating} to allow simulating the motion of torque- and force-free objects advected and rotated by an ambient flow field. These calculations reproduce, in the appropriate limits, approximate Fax\'en relationships developed in Ref. \onlinecite{oppenheimerdiamant2009}, and also capture deviations from approximate predictions when particles are larger than the Saffman-Delbr\"uck length or when the background flow is highly variable. The present approach also allows for a full calculation of the effective viscosity of a rigid-body-laden membrane in the limit of low concentration, i.e. the membrane Einstein correction. Consistent with Ref. \onlinecite{oppenheimerdiamant2009}, cylindrical solid inclusions of radius much smaller than the Saffman-Delbr\"uck length yield an effective membrane viscosity of $\eta_m^\textrm{eff} \approx \eta_m (1 + 2 \phi)$, where $\phi$ is the area fraction of the inclusions. The effective viscosity predictions of Ref. \onlinecite{henle2009effective} have also been confirmed, albeit with the introduction of a rescaling factor of
$2/3$; this same factor appears in comparing Ref. \onlinecite{oppenheimerdiamant2009} to the asymptotic (small particle) limit of Ref. \onlinecite{henle2009effective} and was previously attributed to the different definitions of effective viscosity in these two works. This work shows that this systematic difference is preserved beyond the limit of small particle sizes. The numerical tools used
to confirm these earlier works are immediately transferable to compute effective viscosities of non-circular objects.  For example, {rigid} linear membrane protein chains have increased intrinsic viscosity relative to monomeric membrane proteins, suggesting that membrane protein configurations and oligomerization state could potentially be extracted from measurements of membrane viscosity. The results here provide the numerical tools required to make this possible. However, further study will be required to determine whether experimental measurements of membrane viscosity \cite{camleyesposito2010,petrovschwille_diamonds,petrovschwille2008,honerkamp2013membrane,stebe_microrheology,kim2011interfacial} can be made sufficiently precise to gain useful information from this approach.

The present results may also prove useful in interpreting molecular dynamics simulations of lipid bilayers. Recent molecular simulations have found that increasing the concentration of proteins in the bilayer does increase the effective viscosity of the membrane, but were unable to discriminate between the Oppenheimer-Diamant and Henle-Levine predictions \cite{vogele2019finite}. \fig \ref{fig:oligo} suggests that studying oligomerized proteins may improve the accuracy of these measurements, as the effect of oligomers on membrane viscosity should be larger. 

{Various authors have used
molecular simulations to study the dynamics of membranes crowded with proteins \cite{domanski,javanainen,goose,jeon,javanainen2017diffusion}.  Extending the present intrinsic viscosity calculations to high protein area fractions is nontrivial. Effective viscosity calculations in bulk fluids become complicated outside the dilute limit, due to the competition between hydrodynamic, Brownian, steric, and adhesive effects  \cite{kimkarrila,cichocki1988long,bergenholtz1994huggins,bossis1989rheology}, with phenomenological models commonly used \cite{larson_complex_fluids}. However, the present study highlights a clear physical effect in the dilute regime that should be observable in these detailed simulations and experiments. This may allow for an increasingly quantitative interpretation of dilute systems in terms of effective viscosities. A first step toward understanding complicated crowded membranes may be a careful comparison of theory/numerics, simulations and experiments in dilute membrane systems, where we have a better quantitative understanding of the models.}

\section{ACKNOWLEDGMENTS}
    This work was supported in part by the National Science Foundation (grant No. CHE-1800352).  We thank Haim Diamant and Naomi Oppenheimer for helpful discussions and comments on this work.

\section*{Code Availability}
Code to reproduce all of the figures shown here is available at 
\url{https://github.com/bcamley/Effective-viscosity-reproduce}

\appendix
\section{Deriving effective viscosity}
\label{app:viscosity}
The arguments of Ref. \onlinecite{oppenheimerdiamant2009} are repeated here and slightly generalized to justify the numerical determination of effective viscosity via interfacial RS calculations.

Suppose a point force, $\vb{f}$, is applied to a homogeneous membrane at the origin, leading to a flow field $v^{(0)}_i(\rb) = T^{SD}_{ij}(\rb) f_j$, with $T^{SD}_{ij}$ the Oseen tensor given by Eq. \ref{eq:lubensky}. Now, if the membrane is not homogeneous, but rather includes a rigid object embedded in the membrane at position $\rb'$, that object is subject to a hydrodynamic force dipole, which is compensated by the force dipole by the object acting on the membrane: $S_{ij}$. When averaged over all possible object orientations, this  stresslet is (\eq \ref{eq:stressfaxen})
\begin{equation}
\overline{S}_{ij}(\rb') = -\alpha \eta_m A_p \left[ \partial_i T^{SD}_{jk}(\rb') + \partial_j T^{SD}_{ik}(\rb') \right] f_k 
\label{eq:stresslet_oseen}
\end{equation}
with $\alpha$ determined in practice via the interfacial RS method described in Sec. \ref{sec:effective}.
 $S$ alters the homogeneous flow from $\vb{v}^{(0)}(\rb)$ to $\vb{v}^{(0)}(\rb) + \delta \vb{v}(\rb)$, with $\delta v_i = -S_{kj}(\rb') \partial_k T^{SD}_{ij}(\rb-\rb')$ (this follows from the multipole expansion of the force distribution \cite{kimkarrila}). Averaging over the contributions of $\mathcal{N}$ randomly oriented identical objects, randomly scattered throughout a 
 membrane of area $A_{mem}$, yields
\begin{eqnarray}
\langle \delta v_i(\rb) \rangle &=& -\frac{\mathcal{N}}{A_{mem}}\int_{A_{mem}} d^2 r'  \overline{S}_{kj}(\rb') \partial_k T^{SD}_{ij}(\rb-\rb') \nonumber \\
&=& -\frac{\phi}{A_p}\int_{A_{mem}} d^2 r'  \overline{S}_{kj}(\rb') \partial_k T^{SD}_{ij}(\rb-\rb').
\label{eq:avgdeltav}
\end{eqnarray}
$\phi=\mathcal{N} A_p/A_{mem}$ is the area fraction of particles with $A_p$ the area of each particle.  It should be clear that the above derivation assumes the absence of particle-particle correlations, neglects the contributions of higher order force multipoles and considers only single-particle corrections to $\delta \vb{v}(\rb)$.  Though these approximations are imperfect in general, they are expected to be valid in the small $\phi$ limit assumed in this work.  It is also assumed that the applied force is not exerted for a long enough time for the particles to rearrange in response to the applied flow; previous simulations indicate this is not a significant issue, so long as $\vb{f}$ remains in the linear response regime. \cite{camley2014fluctuating}. 

Fourier transforming \eq \ref{eq:avgdeltav} yields 
\begin{equation}
\langle \delta v_i(\qv) \rangle =  -\frac{\phi}{A_p} \overline{S}_{kj}(\qv) iq_k T^{SD}_{ij}(\qv)
\end{equation}
and Eq. \ref{eq:stresslet_oseen} and \eq \ref{eq:lubensky} imply (note that $q_i T^{SD}_{ij}(\qv) = 0$)
\begin{equation}
    i q_k \overline{S}_{kj}(\qv) = \alpha \eta_m A_p   q^2 T^{SD}_{jm}(\qv) f_m.
\end{equation}
Therefore,
\begin{align}
\langle \delta v_i(\qv) \rangle &= -\phi \alpha \eta_m   q^2 T^{SD}_{ij}(\qv)T^{SD}_{jm}(\qv) f_m \\
						    &= -\alpha\phi \frac{q}{q+\lsd^{-1}} T^{SD}_{im}(\qv) f_m.
\end{align}
The velocity response due to point forcing at the origin and including the effect of an ensemble of randomly
distributed particles is thus
\begin{align}
v_i(\qv) &= v_i^{(0)}(\qv) + \langle \delta v_i(\qv)\rangle \equiv T_{ij}^\textrm{eff}(\qv) f_j \\
T_{ij}^{\textrm{eff}}(\qv) &= \left(1-\alpha \phi \frac{q}{q+\lsd^{-1}} \right) T^{SD}_{ij}(\qv) \\
                            &\approx \frac{1}{\eta_m^{\textrm{eff}} q^2 + 2 \eta_f q } \left(\delta_{ij} - \frac{q_i q_j}{q^2} \right) .
\end{align}
The final line expresses that fact that, to linear order in $\phi$, $T_{ij}^{\textrm{eff}} = T^{SD}_{ij}|_{\eta_m = \eta_m^{\textrm{eff}}}$, with $\eta_m^{\textrm{eff}} = \eta_m \left(1+\alpha \phi\right)$.

\section{Grand Resistance Matrix Formulation}
\label{app:gr}

In earlier work \cite{camley2013diffusion}, the interfacial RS approach was used to compute the forces and torques exerted on rigid bodies embedded in a membrane. In particular, the force $\vb{F}$ and torque $\tau$ required to push a particle with velocity $\vb{U}$ and angular velocity $\Omega$ were determined. The related drag/mobility coefficients were then used to determine translational and rotational diffusion coefficients. Because of the linearity of the Stokes equations, the forces and torques must be {\it linear} in the velocities, 
\begin{equation}
\left( \begin{array}{c} \vb{F}_{hy} \\ \tau_{hy} \end{array} \right) 
= - R \left(\begin{array}{c} \vb{\vb{U}} \\ \Omega \end{array} \right), 
\end{equation}
where $R$ is the $3\times3$ ``resistance matrix".  Here, $\vb{F}_{hy}=-\vb{F}$ and $\vb{\tau}_{hy}=-\vb{\tau}$ are the hydrodynamic
force and torque resisting the external forcing.  The hydrodynamic and external forces (torques) must exactly compensate one another under the steady-state conditions of creeping flow.

A similar linear relationship can be developed for objects embedded in an external flow $\vb{v}^\textrm{amb}$, creating the ``grand resistance matrix" \cite{kimkarrila}, $\mathcal{R}$. If the external flow is slowly-varying enough on the scale of the particle, it can be treated as linear, and broken down as $\vb{v}^\textrm{amb}(\rb) \approx \vb{U}^\infty + \Omega^\infty \times \rb + E^\infty \cdot \rb$, where the linear flow has been broken into a rotational term and a straining flow. 

While a rigid object can rotate or translate in response to an externally imposed flow, it cannot deform. In resisting an attempted deformation by the surrounding fluid, it will exert forces on the fluid; at lowest order in the multipole expansion, this generates a net force dipole with strength $S_{ij}=-S^{hy}_{ij}$ exactly compensating the hydrodynamic dipole imposed by the external flow on the object. The force, torque, and force dipole exerted on an object will thus be given by a linear relationship of the form\cite{kimkarrila}
\begin{align}
\left( \begin{array}{c} \vb{F}_{hy} \\ \tau_{hy} \\ S_{hy} \end{array} \right)  &=  \mathcal{R} \left(\begin{array}{c} \vb{U}^\infty-\vb{U} \\ \Omega^\infty-\Omega \\ E^\infty \end{array} \right)  \nonumber \\
&= 
\left(\begin{array}{ccc} A & \widetilde{B} & \widetilde{G} \\ B & C & \widetilde{H} \\ G & H & M \end{array}\right) \left(\begin{array}{c} \vb{U}^\infty-\vb{U} \\ \Omega^\infty-\Omega \\ E^\infty \end{array} \right) \label{eq:gr}
\end{align}
Eq. \ref{eq:gr} introduces a simplifying notation: blocks of the grand resistance matrix are scalars (e.g. $C$), vectors (e.g. $B$), and higher-order tensors (e.g. $A$ or $M$). Multiplication implies contraction across the indices, i.e. \eq \ref{eq:gr} states that if $\vb{U}=\vb{U}^\infty$ and $\Omega = \Omega^\infty$, $S = M E^\infty$, which is to be taken as $S_{ij} = M_{ijkl}E^\infty_{kl}$. In addition, there are important symmetry relationships in these terms that follow from the Reciprocal Theorem \cite{kimkarrila}. In the quasi-two-dimensional system under study, where the only nonzero components of the angular velocity and torque are in the $z$ direction, these relationships are:
\begin{align}
A_{ij} &= A_{ji}  \nonumber  \\
B_{i} &= \widetilde{B}_i \nonumber \\
G_{ijk} &= \widetilde{G}_{kij} \nonumber \\
H_{ij} &= \widetilde{H}_{ij} \nonumber \\
M_{ijkl} &= M_{klij} \label{eq:symmetry}
\end{align}
where there is no symmetry relationship for $C$ because it is a scalar.  The grand resistance matrix can be used to simply compute the effective viscosity of a rigid body-laden membrane, as will be shown below. 

The grand resistance matrix $\mathcal{R}$ can be calculated as a straightforward extension of the results of Ref. \onlinecite{camley2013diffusion}. The basic approach is to choose the velocity of the blob points to be consistent with a rigid body motion, find the forces $g_j[\vb{R}_n]$ that generate these velocities when subjected to $\vb{v}^\textrm{amb}(\rb) =  \vb{U}^\infty + \Omega^\infty \times \rb + E^\infty \cdot \rb$ , and then use the blob forces to compute $\vb{F}$,$\tau$, and $S_{ij}$, which allow one to read off the matrix elements of \eq \ref{eq:gr}. For instance, to find the (translational) drag coefficient matrix $A_{ij}$, start with \eq \ref{eq:reg} and choose $v_i(R_m) = U_{i}$ and $\vb{v}^\textrm{amb} = 0$. The blob forces $\vb{g}\left[\vb{R}_n\right]$ that generate this motion are determined numerically and are used to compute the total force on the particle $\vb{F} = \sum_n \vb{g}\left[\vb{R}_n\right]$ that drives the translational motion. The hydrodynamic drag force $\vb{F}_{hy} = -\vb{F}$. Then, from \eq \ref{eq:gr}, read off $\vb{F}_{hy} = -\sum_{n} \vb{g}\left[\vb{R}_n\right] = A(\vb{U}^\infty-\vb{U}) = -A\vb{U}$. For the choice $\vb{U} = \hat{\vb{x}}$,  $A_{xx} = \sum_{n} g_x\left[\vb{R}_n\right]$ and $A_{yx} = \sum_{n} g_y\left[\vb{R}_n\right]$. Choosing $\vb{U} = \hat{\vb{y}}$ allows the reconstruction of the remaining components of $A$. This process can be extended straightforwardly for nearly all of the components of $\mathcal{R}$.  Furthermore, the symmetry relationships of \eq \ref{eq:symmetry} obviate the need to perform all the calculations that one might naively expect were needed. 

There is one subtlety in constructing $M_{ijkl}$. To access $M_{ijkl}$, one would like to choose $\vb{U}=0$, $\Omega = 0$, and $v^\textrm{amb} = -E_{ij} r_j$. For this choice, Eq. \ref{eq:gr} suggests that $S^{hy}_{ij} = -S_{ij} = -\frac{1}{2} \sum_n \left(R_{n,i} g_j[\vb{R}_n] + R_{n,j} g_i[\vb{R}_n] \right) = -M_{ijkl}E_{kl}$. In principle, then, it should be possible to choose $E_{kl} = 1$ for one component $kl$, e.g. $kl = xy$, and zero otherwise. Then it would be possible to determine $M_{ijxy} = \frac{1}{2} \sum_n \left(R_{n,i} g_j[\vb{R}_n] + R_{n,j} g_i[\vb{R}_n] \right)$, and then repeat this computation for each component. The difficulty in doing this is that $E_{ij}$ must obey two constraints: 1) $E_{ij} = E_{ji}$, and 2) for incompressible flow, $\nabla \cdot \vb{v}^\textrm{amb} = 0$ and $\textrm{tr} E = E_{ii} = 0$. Thus there are only two independent components of $E$, and any physical $E$ can be built out of $E = \left(\begin{array}{cc} 1 & 0 \\ 0 & -1 \end{array} \right)$ and $E = \left(\begin{array}{cc} 0 & 1 \\ 1 & 0 \end{array} \right)$. 

Choosing $v^\textrm{amb} = -E_{ij} r_j$ with $$E = \left(\begin{array}{cc} 1 & 0 \\ 0 & -1 \end{array} \right)$$ then shows $M_{ijxx}-M_{ijyy} = \frac{1}{2} \sum_n \left(R_{n,i} g_j[\vb{R}_n] + R_{n,j} g_i[\vb{R}_n] \right)$, where the forces are the ones required to establish this particular flow. Only the combination $M_{ijxx}-M_{ijyy}$ can be identified. However, for any physical flow field $E^\infty$, $S_{hy} = M E^\infty$ will only depend on $M_{ijxx}$ and $M_{ijyy}$ in the combination $M_{ijxx}-M_{ijyy}$, because $E^\infty$ must have zero trace. We therefore choose to set $M_{ijxx}+M_{ijyy} = 0$ to uniquely specify $M_{ijxx}$ and $M_{ijyy}$, though any arbitrary constant would yield the same physical results.

Similarly,  $v^\textrm{amb} = -E_{ij} r_j$ with $E = \left(\begin{array}{cc} 0 & 1 \\ 1 & 0 \end{array} \right),$ yields $M_{ijxy}+M_{ijyx} = \frac{1}{2} \sum_n \left(R_{n,i} g_j[\vb{R}_n] + R_{n,j} g_i[\vb{R}_n] \right)$ -- note again that $g_j$ are the forces for this flow. In this case, the combination $M_{ijxy}-M_{ijyx}$ is not determined, but is also not physically relevant. We arbitrarily choose the convention $M_{ijxy}=M_{ijyx}$. 

\subsection*{Computing effective viscosity with the grand resistance matrix}

Once the grand resistance matrix is known, it can be used to compute the effective viscosity. 
A particle with no external force or torque exerted on it, but subject to hydrodynamic perturbations from the surrounding fluid, will exert a force dipole $S_{ij}$ on the membrane. How does this depend on the external flow $(\vb{U}^\infty,\Omega^\infty,E_{ij}^\infty)$? The answer is hidden inside \eq \ref{eq:gr}. In the absence of any external force or torque on the particle, Eq. \ref{eq:gr} suggests
\begin{align}
\left( \begin{array}{c} \vb{F} \\ \tau \end{array} \right) =
\left( \begin{array}{c} \vb{F}_{hy} \\ \tau_{hy} \end{array} \right)
   &= \left( \begin{array}{c} \mathbf{0} \\ 0 \end{array} \right)  \\
   &= 
\left(\begin{array}{ccc} A & \widetilde{B} & \widetilde{G} \\ B & C & \widetilde{H} \end{array}\right) \left(\begin{array}{c} \vb{U}^\infty-\vb{U} \\ \Omega^\infty-\Omega \\ E^\infty \end{array} \right) 
\end{align}
with the solution
\begin{equation}
    \left(\begin{array}{c}\vb{U}-\vb{U}^\infty \\ \Omega-\Omega^\infty \end{array}\right) = \left(\begin{array}{cc} A & \widetilde{B}  \\ B & C  \end{array}\right)^{-1}\left( \begin{array}{c} \widetilde{G} \\ \widetilde{H} \end{array} \right)E^\infty.
\end{equation}
Upon substitution of this result into Eq. \ref{eq:gr}, an expression for
the hydrodynamic dipole is obtained
\begin{align}
\nonumber S_{hy} &= -S \\&=  \left [ M - \left( G H\right)\left(\begin{array}{cc} A & \widetilde{B}  \\ B & C  \end{array}\right)^{-1}\left( \begin{array}{c} \widetilde{G} \\ \widetilde{H} \end{array} \right)\right ]E^\infty, \label{eq:correction} \\
&\equiv \mathcal{M} E^\infty
\end{align}
which is the quasi-2D analog to Eq. 5.21 of Ref. \onlinecite{kimkarrila}.
The rank-4 tensor in square brackets in \eq \ref{eq:correction} is the sum of two terms.  The
first term ($M$) indicates the dipolar response to attempted straining deformations by the fluid for a particle translating and rotating with the background flow.  The second term corrects this response to impose conditions of vanishing external force/torque.  By virtue of the symmetry relations on the resistance matrix (Eq. \ref{eq:symmetry}), this ``correction term" is obtainable entirely from traditional RS calculations in the absence of $E^{\infty}$.  $M$ itself (up to the physically unimportant constants mentioned above) is obtainable from RS calculations for a static object $\vb{U}=\Omega=0$ with imposed $E^\infty$.

Appendix \ref{app:viscosity} shows that, if the force dipole obeys
\begin{equation}
S_{ij} \approx -\alpha \eta_m A_p \left( \partial_i v_{j}^\textrm{amb}+\partial_j v_{i}^\textrm{amb} \right), \label{eq:appB}
\end{equation}
then $\alpha$ is the intrinsic viscosity.  This conclusion assumes slowly-varying external velocities. When flow is slowly-varying enough to allow approximation as $\vb{v}^\textrm{amb}(\rb) \approx \vb{U}^\infty + \Omega^\infty \times \rb + E^\infty \cdot \rb$, it is easily seen that Eq. \ref{eq:appB} reduces to
\begin{equation}
S_{ij} \approx -\alpha \eta_m A_p \left( E^\infty_{ij} + E^\infty_{ji} \right).
\end{equation}
As noted previously, \eq \ref{eq:appB} holds only for {\it isotropic} particles; orientational averaging is required to yield this result in the general case. 

$\alpha$ is calculated by picking a background flow field $E^\infty$, then computing $S_{ij}$ induced by the particle in that flow field directly from $\mathcal{R}$ by using \eq \ref{eq:correction}. This is repeated for different particle orientations and values of the blob spacing and $\epsilon$, averaging over orientations, and extrapolating to zero spacing.  $\alpha$ can be computed as
\begin{equation}
    \alpha = \frac{-\overline{S}_{xx}}{2 \eta_m A_p}
\end{equation}
for the choice $E^{\infty} = \left(\begin{array}{cc} 1 & 0 \\ 0 & -1 \end{array} \right)$. 

In fact, because \eq \ref{eq:correction} determines the full tensor $\mathcal{M}$ that generates the hydrodynamic dipole response, orientational averaging may be carried out analytically, significantly improving computational speed. This resembles the approach of Ref.  \onlinecite{andrews1977three}, but in two dimensions. When the particle is rotated, the tensor $\mathcal{M}_{ijkl}$ transforms to $R_{ai}R_{bj}R_{ck}R_{dl} \mathcal{M}_{ijkl}$, where $R_{ai}$ are the appropriate rotation matrices. The average over the angle of rotation follows as
\begin{equation}
Z_{abcdijkl} \equiv \frac{1}{2\pi}\int_0^{2\pi} d\theta R_{ai}R_{bj}R_{ck}R_{dl}.
\end{equation}
These integrals are simple, if tedious, to evaluate -- they are products of sines and cosines, which may be automated in Mathematica or constructed following Ref. \onlinecite{andrews1977three}. Then, the orientationally-averaged force dipole is given by
\begin{equation}
    \overline{S}_{ab} = -Z_{abcdijkl}\Mm_{ijkl} E^\infty_{cd}
\end{equation}
For present purposes, only $\overline{S}_{xx}$ is needed, which is given by
\begin{align}
\nonumber \overline{S}_{xx} = -\frac{1}{4}\left(\Mm_{xxxx} - \Mm_{xxyy} + \Mm_{xyxy} + \Mm_{xyyx} \right. \\
\left. +  \Mm_{yxxy} + \Mm_{yxyx} - \Mm_{yyxx} + \Mm_{yyyy}\right) \nonumber
\end{align}
where $\Mm$ here is the tensor evaluated from \eq \ref{eq:correction} {\it at a single orientation} and we have assumed $E^{\infty} = \left(\begin{array}{cc} 1 & 0 \\ 0 & -1 \end{array} \right)$. 

This approach has been used to reproduce the computations in \fig \ref{fig:oligo}.  The maximum relative difference
between the two approaches is less than $2 \times 10^{-5}$.  Indeed, the two calculations should be completely
equivalent, up to the differences between numerical versus analytical orientational averaging and the accuracy 
of the numerical method.

\onecolumngrid
\section*{References}

\end{document}